\documentclass{PoS}

\def\nch{\ensuremath{n_{\mathrm{ch}}}}
\def\Nch{\ensuremath{N_{\mathrm{ch}}}}
\def\meanpT{\ensuremath{\langle p_\mathrm{T} \rangle}}
\def\pT{\ensuremath{p_\mathrm{T}}}

\title{Highlights from ATLAS}

\ShortTitle{Highlights from ATLAS}

\author{\speaker{Thorsten Wengler}%
        \thanks{Now CERN, Geneva, Switzerland}\\
       University of Manchester\\
       For the ATLAS Collaboration\\
       E-mail: \email{Thorsten.Wengler@cern.ch}}

\abstract{The ATLAS experiment has been taking data efficiently since LHC collisions started, first at the injection energy of 450~GeV/beam and at 1.18~TeV/beam in 2009, then at 3.5~TeV/beam in 2010. Many results have already been obtained based on this data demonstrating the performance of the detector, as well as first physics measurements. Only a selection of highlights will be presented here. }

\FullConference{XVIII International Workshop on Deep-Inelastic Scattering and Related Subjects\\
		 April 19 -23, 2010\\
		 Convitto della Calza, Firenze, Italy}

\begin{document}

\section{Introduction}

ATLAS, one of the large collider detectors built into the collision regions of the Large Hadron Collider (LHC) at CERN, has started collecting data from proton-proton collisions from the first day they were provided by the accelerator in November 2009. After many years of commissioning the detector with cosmic ray data, and more recently with events related to single beams passing the detector, the performance of each component could be verified and refined based on collision events, while at the same time extracting the very first physics results.

The scope of this article allows for only a fraction of the results obtained to be discussed. Detector components are described in detail in~\cite{Aad:2008zzm}. They will be introduced briefly and their performance will be described. The detector is in excellent condition, with 98\% or more of signal channels operational for all sub-systems. Calibration and performance studies are progressing rapidly, in some cases already approaching nominal detector performance.

The data sample from the 2009 data taking consists of 917k events (538k events during stable beam operations with all detector components switched on), corresponding to an integrated luminosity of 20~$\mu$b$^{-1}$ (12~$\mu$b$^{-1}$ during stable beam operation). The peak luminosity reached in 2009 was $\sim 7 \times 10^{26}$~cm$^{-2}$s$^{-1}$. Most of the data was taken at the injection energy of $\sqrt{s}=900$\,GeV, with 34k events taken at $\sqrt{s}=2.36$\,TeV. In 2010 ATLAS only considered data taken during stable beam periods for further analysis. At the time of the conference, the data sample consisted of $17.8 \times 10^6$ events, corresponding to an integrated luminosity of 307~$\mu$b$^{-1}$, with a peak luminosity $\sim 1.9 \times 10^{27}$~cm$^{-2}$s$^{-1}$. These numbers have since been surpassed by several orders of magnitude.

\section{Trigger and data acquisition}

ATLAS uses a three-stage trigger system to filter interesting events from the up to 40\,MHz collision rate that can be provided by the LHC. The first stage uses custom-built electronics to reduce the rate to a maximum of currently 50\,kHz (eventually up to 100\,kHz), using fast information provided by the calorimeters and the muon spectrometer. In addition to the trigger accept signal, the first level trigger also provides geometrical information to the higher trigger levels, indicating 'Regions-of-Interest (RoI)' where signals fulfilling trigger requirements have been found. Based on these inputs and further readout from the detector, the High-Level-Trigger (HLT) reduces the rate in two further steps. The Level-2 trigger complements the RoI information with partial event readout to achieve an output rate of not more than $\sim 3$\,kHz, while the Event Filter (EF) uses fully built events and sophisticated reconstruction algorithms to achieve an average output rate to disk of $\sim 200$\,Hz, depending on running conditions and the experimental setup. Both Level-2 and EF are implemented as software algorithms running on a farm of several thousand CPUs.

By the time of the conference, event selection was still performed by the Level-1 trigger alone, as the still low event rate did not require active selection by the HLT. However, many HLT algorithms were already run in 'monitoring mode', which executes the trigger chains without making use of their decisions for event selection. This not only provides excellent testing of the HLT stability and performance, but also produces alternative trigger information for efficiency studies and calibration information like the online beam spot determination.

\begin{figure}
\includegraphics[angle=90,width=1.0\textwidth]{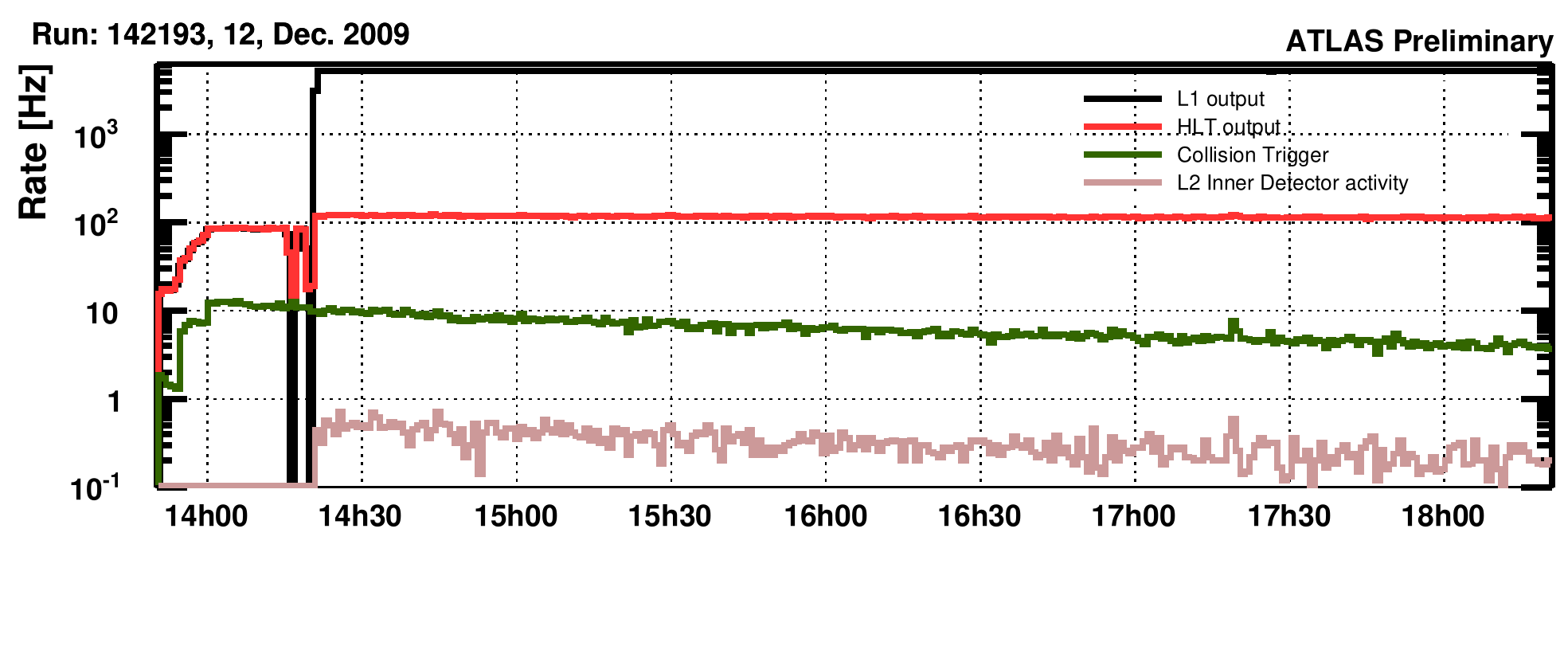}
\caption{L1 and HLT trigger rates for a typical run with stable beams. Also shown are a collision trigger at L1, requiring hits on both the A and the C side of the minimum bias scintillator counters and filled bunches for both beams. The line labelled L2 Inner Detector activity represents a filtering algorithm at the L2 trigger, which accepts events based on space point counts in the inner detector.}
\label{fig-hltrej}
\end{figure}

Figure~\ref{fig-hltrej} shows the L1 and HLT trigger rates for a typical run with stable beams. Also shown are a collision trigger at L1, requiring hits on both the A and the C side of the minimum bias scintillator counters and filled bunches for both beams. The line labelled L2 Inner Detector activity represents a filtering algorithm at the L2 trigger, which accepts events based on space point counts in the inner detector. This L2 algorithm receives 5\% of all filled bunches as input from L1. Assuming both the L1 collision trigger and the space point counting are highly efficient for collision events, the difference in the two lines should reflect this fraction, even though the acceptance of both triggers is different. The moment the L2 algorithm is enabled is clearly visible as the jump of output L1 rate, and the start of event rate on the L2 line. The dips in HLT and L1 output rates just before this moment are due to the short pause needed to change trigger setup. The HLT output rate (which represents the rate of events recorded to disc) does not visibly change, as it is dominated by a constant rate of monitor triggers.

\begin{figure}
\includegraphics[width=0.60\textwidth]{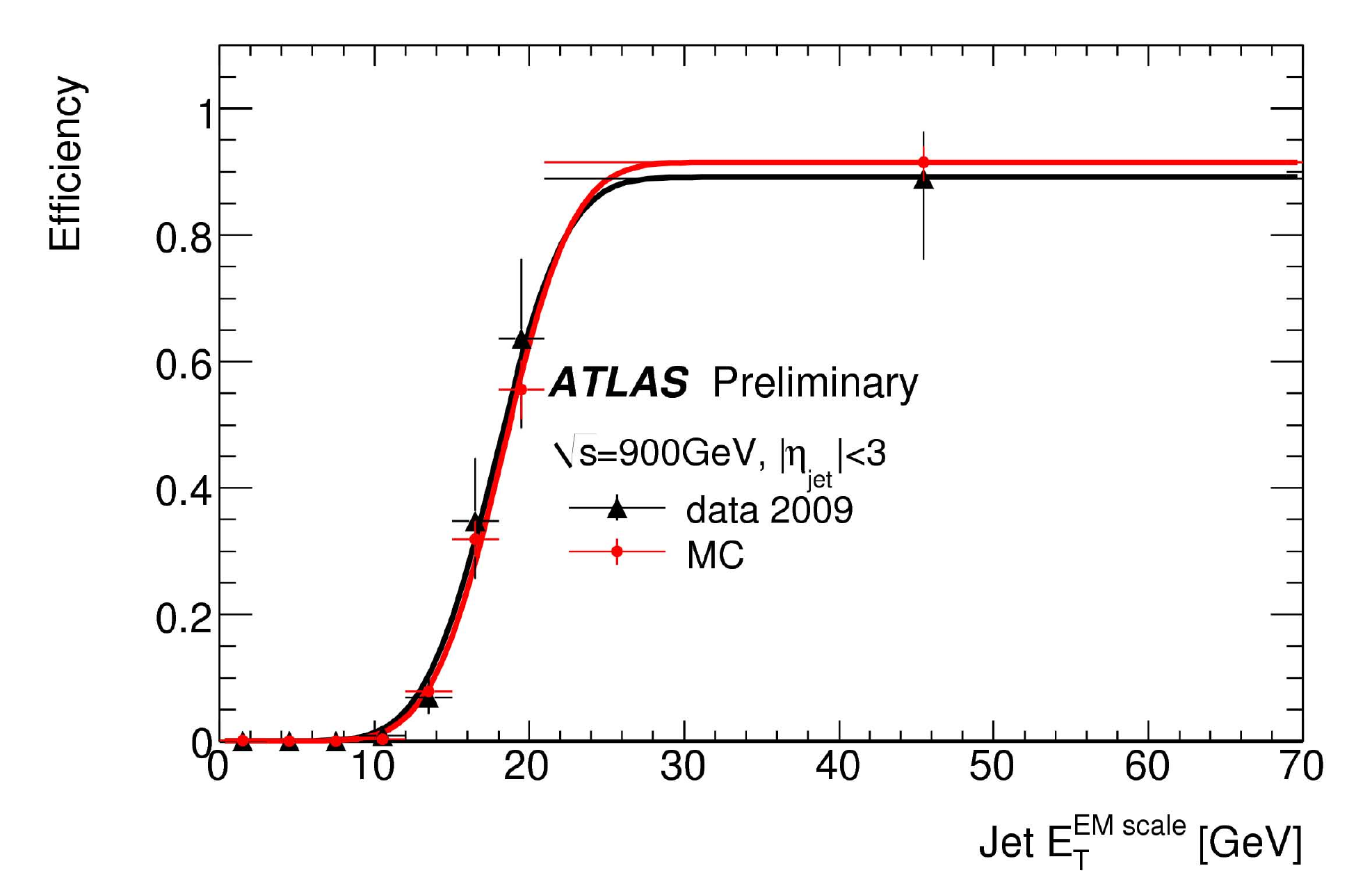}
\includegraphics[width=0.40\textwidth]{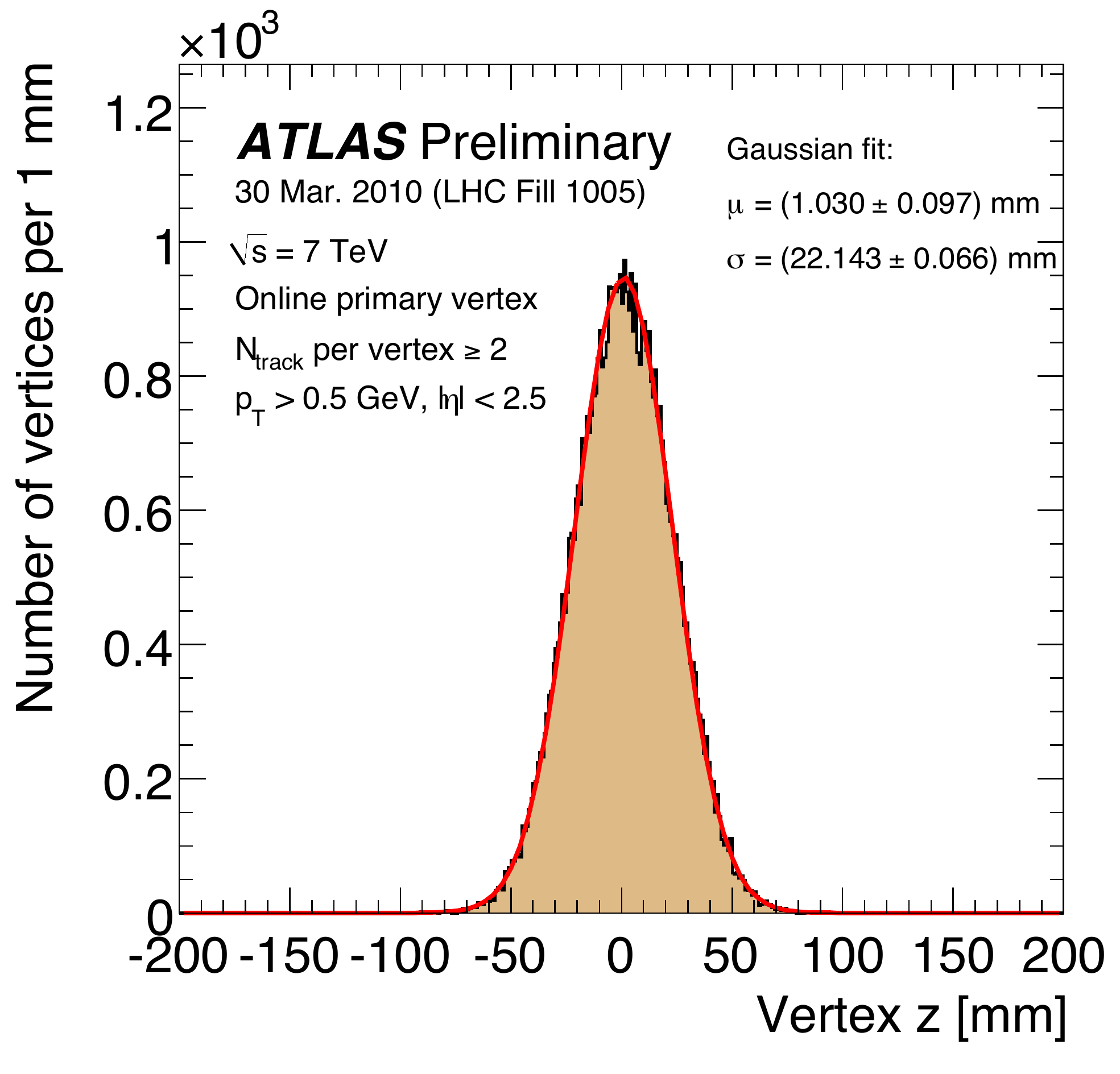}
\caption{Left: L1 Efficiency for the trigger selecting jets above 10 counts ($\sim 10$\,GeV) as a function of the offline jet transverse momentum at the electromagnetic energy scale. The turn-on is shown for data (black triangle points) and Monte Carlo simulations (red circle points).
Right: Online longitudinal luminous centroid (longitudinal "beam spot" position) for a run at $\sqrt{s}=7$\,TeV. }
\label{fig-jet+vtx}
\end{figure}

Figure~\ref{fig-jet+vtx} (right) shows the longitudinal luminous centroid (longitudinal "beam spot" position) at $\sqrt{s}=7$\,TeV. These measurements were available online in real time as soon as the HLT was activated for the first high energy collisions at the LHC. Gaussian fits (within $\pm1\times RMS$ for $x$,$y$) are used to extract the luminous region mean position and width, where the latter is dominated by the vertexing resolution. An excellent agreement is observed among different tracking algorithms online, and with respect to the more sophisticated offline beam spot measurement. While the event selection at this stage relied almost exclusively on the inclusive and efficient minimum bias collision trigger, many other trigger signatures where already active as well, with all trigger decisions recorded in the data written to disk. Figure~\ref{fig-jet+vtx} (left) shows the first level trigger efficiency for the trigger selecting jets above 10 counts ($\sim 10$\,GeV) as a function of the offline jet transverse momentum at the electromagnetic energy scale. The turn-on is shown for $\sqrt{s}=900$\,GeV data (black triangle points) and Monte Carlo simulations (red circle points). The turn-on curve shows the expected behaviour. There is good agreement between the data and the Monte Carlo simulation.

The ATLAS trigger and data acquisition system has performed very well during the startup of the experiment, and following the changing machine conditions every since, with necessary adjustments to the system done with little or no interference with the data taking. The data taking efficiency for periods of stable beam is above 90\%, an excellent achievement especially given the early stage of the data taking.

\section{Inner detector}

The ATLAS inner tracking detector consists of three distinct sub-detectors. A silicon pixel detector closest to the beam pipe is followed by a silicon strip detector (SCT), which in turn is surrounded by the transition radiation tracker (TRT), which employs gas-filled 'straws' with a central wire and radiation foils to provide both tracking and particle identification by transition radiation. The inner detector is embedded in a 2\,T solenoidal magnetic field.

\begin{figure}
\includegraphics[width=0.47\textwidth]{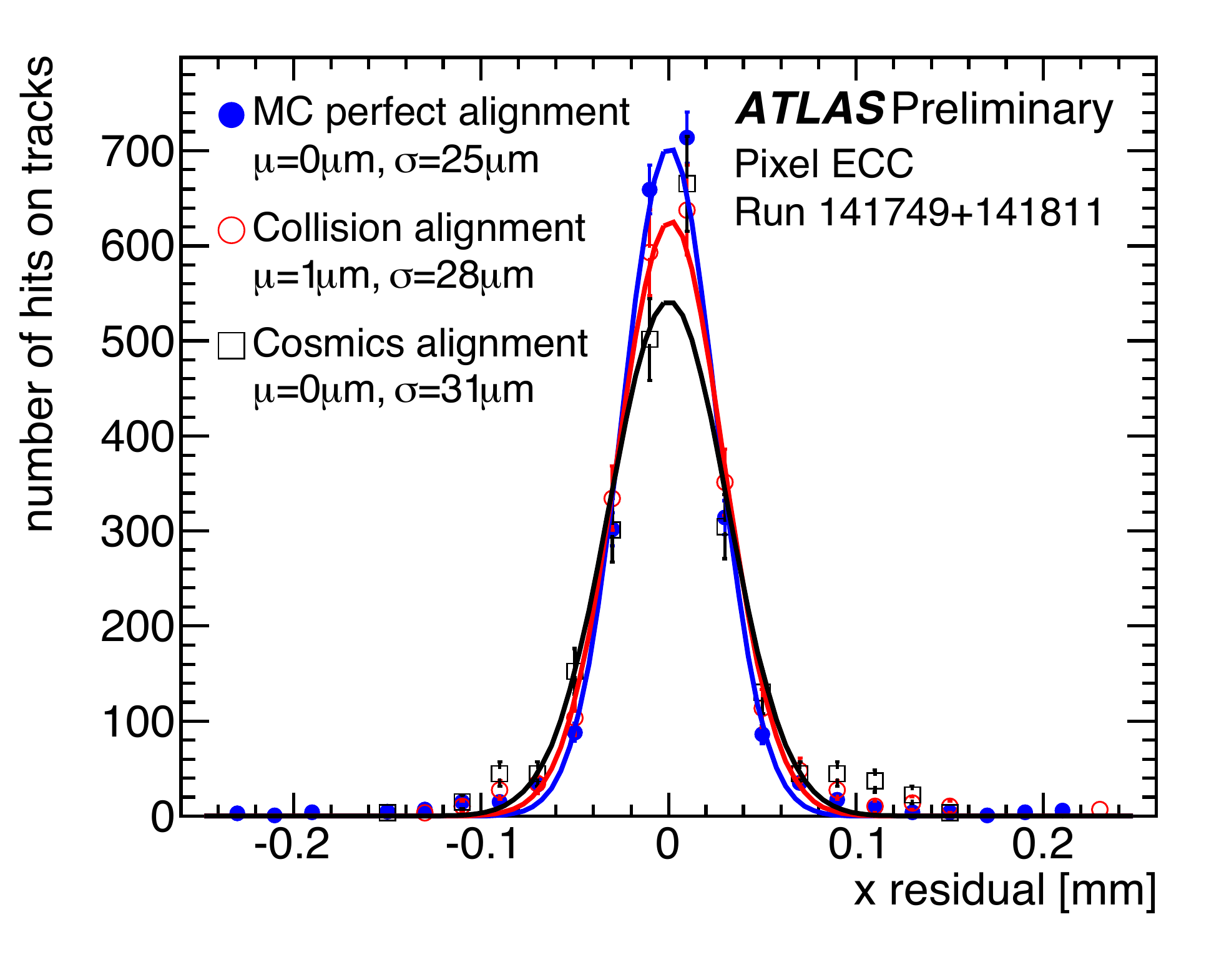}
\includegraphics[width=0.53\textwidth]{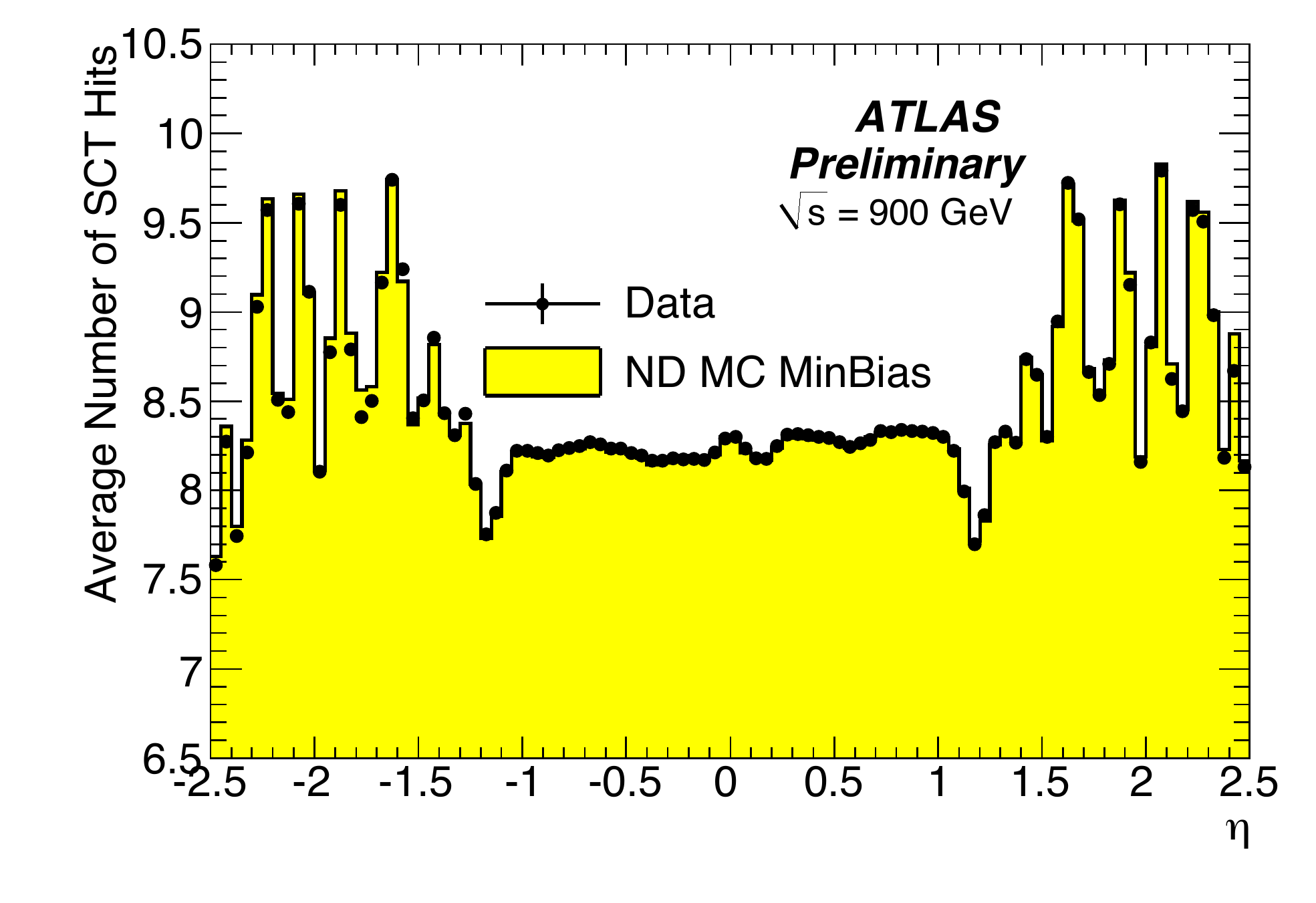}
\caption{Left: Pixel end-cap side C local x residuals for a minimum bias Monte Carlo sample with perfect alignment (solid blue), collision data using the alignment based on cosmic rays (open black squares) and after a first update using collision data (open red circles). A single Gaussian fit is performed.
Right: Comparison between number of SCT hits on reconstructed tracks in $\sqrt{s}=900$\,GeV data and non-diffractive minimum bias Monte Carlo simulation. The plot shows the $\eta$ distribution in which the increase in the number of hits in the end cap region is clearly visible. The structure of the SCT disks is reproduced by the simulation.  }
\label{fig-pixsct}
\end{figure}

\begin{figure}
\includegraphics[width=0.52\textwidth]{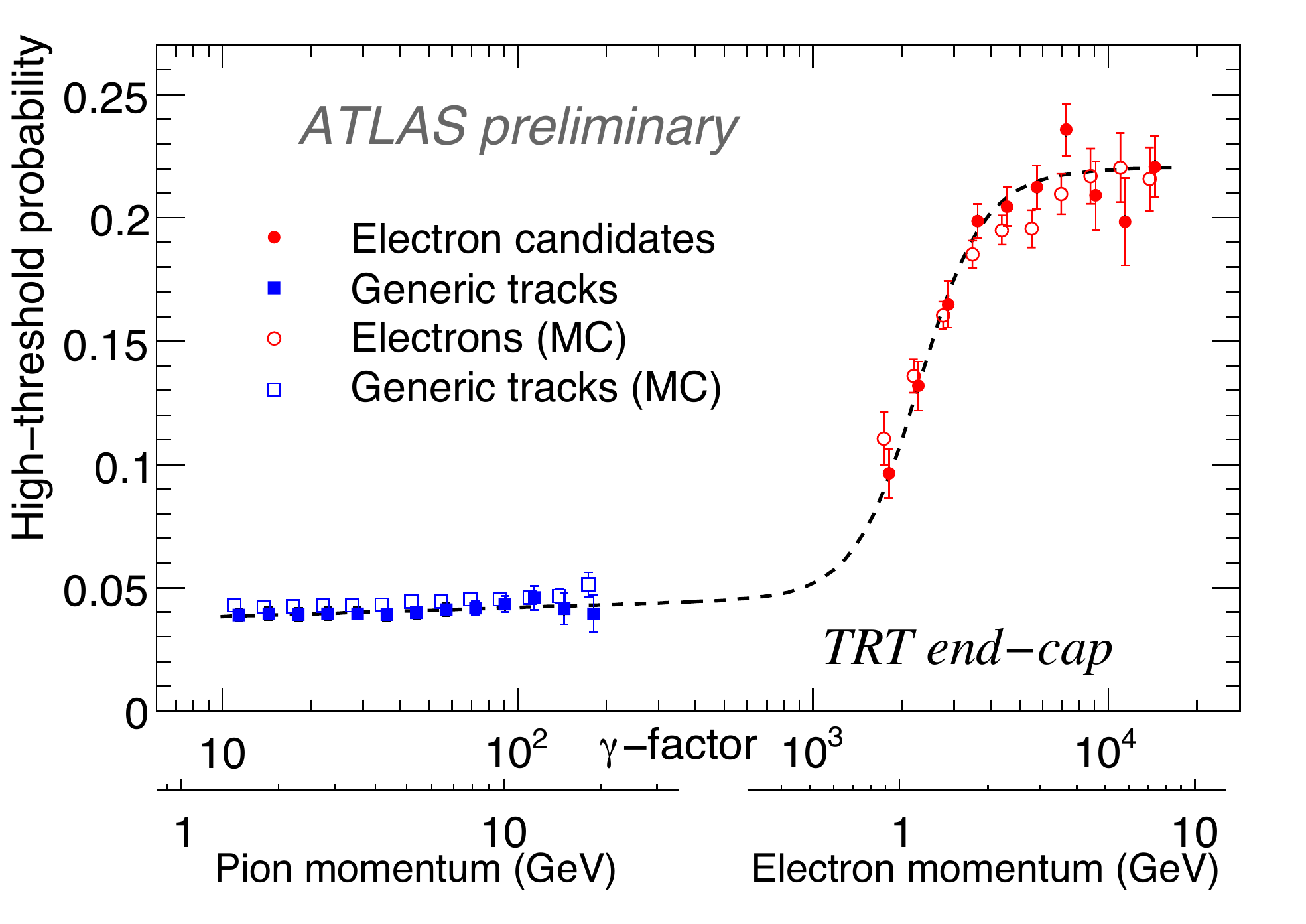}
\includegraphics[width=0.48\textwidth]{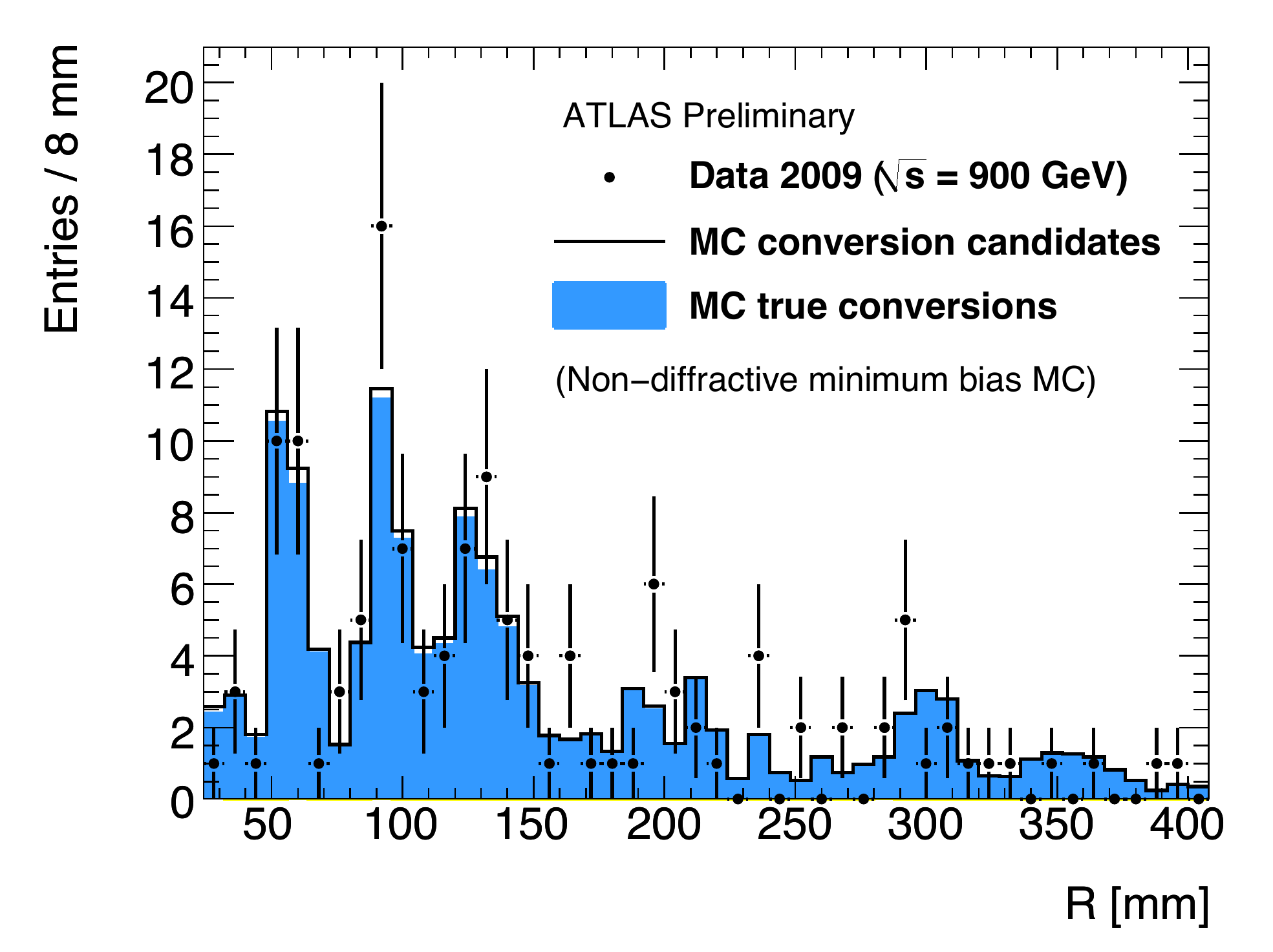}
\caption{Left: The plot shows the probability of a TRT high-threshold hit as a function of the Lorentz factor $\gamma$ = E/m for the TRT end-caps, as measured in LHC collision events. The onset of the production of transition radiation for particles with a $\gamma$-factor above 1000 can be seen. 
Right: Distribution of photon conversion candidate radius for $|\eta|>1.4$. The points show the distribution for all photon conversion candidates in data. The corresponding distribution from the Monte Carlo simulation are shown as the open histogram and the filled histogram shows the contribution of true photon conversions as predicted from the Monte Carlo simulation.}
\label{fig-trtconv}
\end{figure}

Starting with data collected during the cosmic ray data taking, and now using collision events, excellent results have already been achieved in aligning the detectors. Figure~\ref{fig-pixsct} (left) shows the track residuals in the x-coordinate for one of the two pixel detector end-caps. The alignment initially obtained from cosmic ray data is compared to the refined values as obtained from collision data and to the ideal situation, represented by the Monte Carlo simulation. It is evident that even after the short period of data taking completed by the time of the conference, the alignment is already close to the nominal performance. The same holds for the other components of the inner detector. Figure~\ref{fig-pixsct} (right) compares the number of hits recorded in the SCT for reconstructed tracks in $\sqrt{s}=900$\,GeV data and Monte Carlo simulation. This distribution is very sensitive to the correct modelling of the detector geometry and material. The simulation reproduces the structure of the detector very well. Figure~\ref{fig-trtconv} (left) shows the probability of a TRT high-threshold hit as a function of the Lorentz factor $\gamma$ = E/m for the TRT end-caps, as measured in LHC collision events.
The onset of the production of transition radiation for particles with a $\gamma$-factor above 1000 can be seen. By detecting the transition radiation photons, which deposit additional 
energy, thus producing higher-amplitude signals, which trigger the high-threshold discriminator in the front-end electronics, the TRT is able to separate electrons from pions over the momentum range between 1\,GeV and 150\,GeV. To demonstrate this feature, the high-threshold hit probabilities of two samples are compared as a function of their $\gamma$-factor:
\begin{itemize}

\item Electron candidates (full red circles: data, open red circles: non-diffractive minimum bias Monte Carlo simulation) are selected from photon conversions, which have a good-quality mass-constrained reconstructed vertex at a radius greater than 60\,mm. When one track is identified as an electron, the other track is considered as an electron candidate. The purity of the sample is considerably improved by applying tight electron identification cuts on one of the tracks, while the second track is considered as the electron candidate providing an essentially unbiased measurement of the transition radiation performance. The sample includes 855 tracks;

\item Tracks to which no selection criterion was applied (generic tracks, filled blue squares: data, open blue squares: Monte Carlo) are assumed to be pions, 1.3 million tracks are in this sample.

\end{itemize}
The dashed line is an illustrative fit to the data points and indicates the onset of the production of transition radiation as it is expected for the TRT end-cap radiators. The systematic uncertainty from possible backgrounds and selection biases is of the same order as the statistical uncertainty.
The onset of the transition radiation is measured for the first time using data with electron candidates from photon conversions. This provides the TRT detector with an excellent starting point to study and optimise its particle identification properties.

Figure~\ref{fig-trtconv} (right) shows the distribution of radii of photon conversion candidates for $|\eta|>1.4$. The points show the distribution for all photon conversion candidates in data; the open histogram, the corresponding distributions from the Monte Carlo simulation and the filled histogram shows the contribution of true photon conversions as predicted from the Monte Carlo simulation. The radial distributions is shown for $R>24$\,mm, which corresponds to about six times the resolution of the vertex radial position, to ensure a good reconstruction of the $\eta$ of the vertex. The distributions are normalised to the same number of photon conversion candidates in data and Monte Carlo simulation. The structures in the distribution correspond to the material distribution in the detector, very visible are for example the three layers of the Pixel detector between 50\,mm and 150\,mm radius. With more statistics this type of distribution will provide a stringent test of the material description in the simulation.

\begin{figure}
\includegraphics[width=0.50\textwidth]{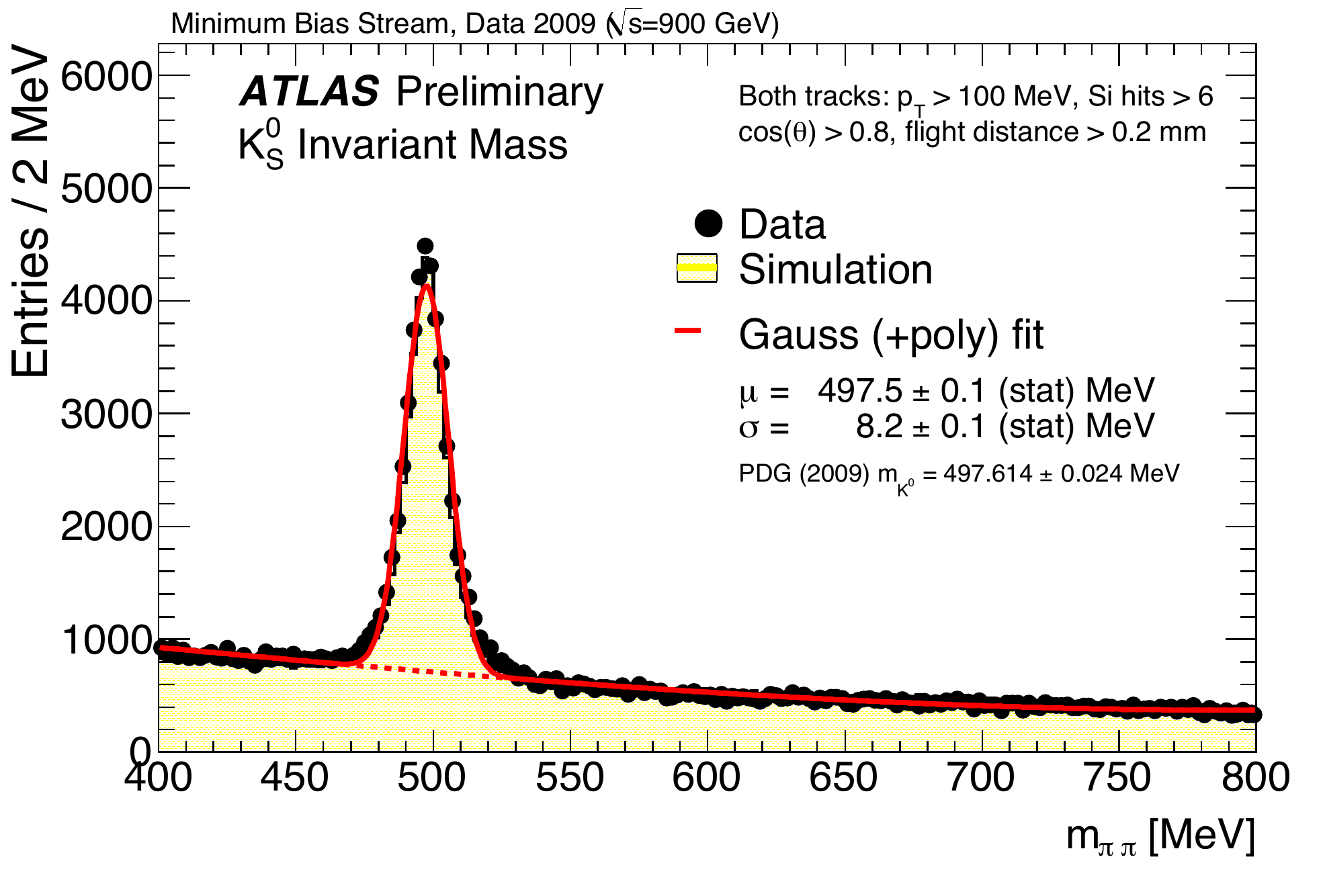}
\includegraphics[width=0.50\textwidth]{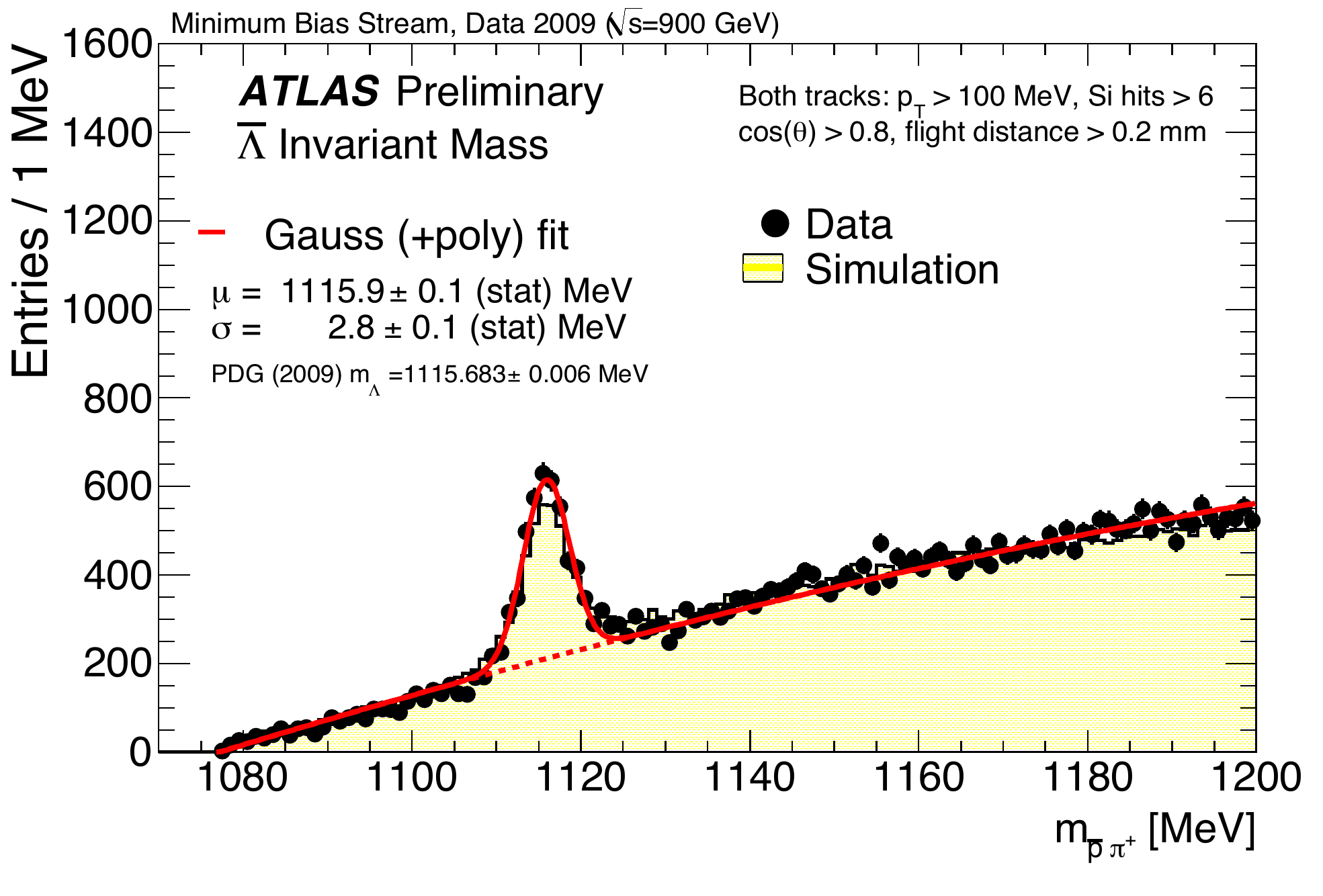}
\caption{Left: The invariant mass distribution of two track vertices found with the ATLAS standard vertex finding code in the invariant mass range range 400 to 800 MeV. No mass constraint is applied during the vertex fit. 
Right: Identical vertex selection in the invariant mass range 1000 to 1200 MeV.  }
\label{fig-kslam}
\end{figure}

A powerful way to optimise and verify the performance of the tracking detectors is to reconstruct resonances with well established mass and decay width. The ATLAS collaboration has carried out many such studies. Two examples shown here. Figure~\ref{fig-kslam} (left) shows the invariant mass distribution of two track vertices found with the ATLAS standard vertex finding code in the invariant mass range 400 to 800\,MeV. No mass constraint is applied during the vertex fit. The K$^\mathrm{0}_\mathrm{S}$ mass peak is clearly visible and well described by the simulation. The same vertex selection in the invariant mass range from 1000 to 1200\,MeV yields the mass peak corresponding to the $\bar{\Lambda}$, again well reproduced by the simulation.

\section{Calorimetry}

\begin{figure}[b]
\includegraphics[width=0.50\textwidth]{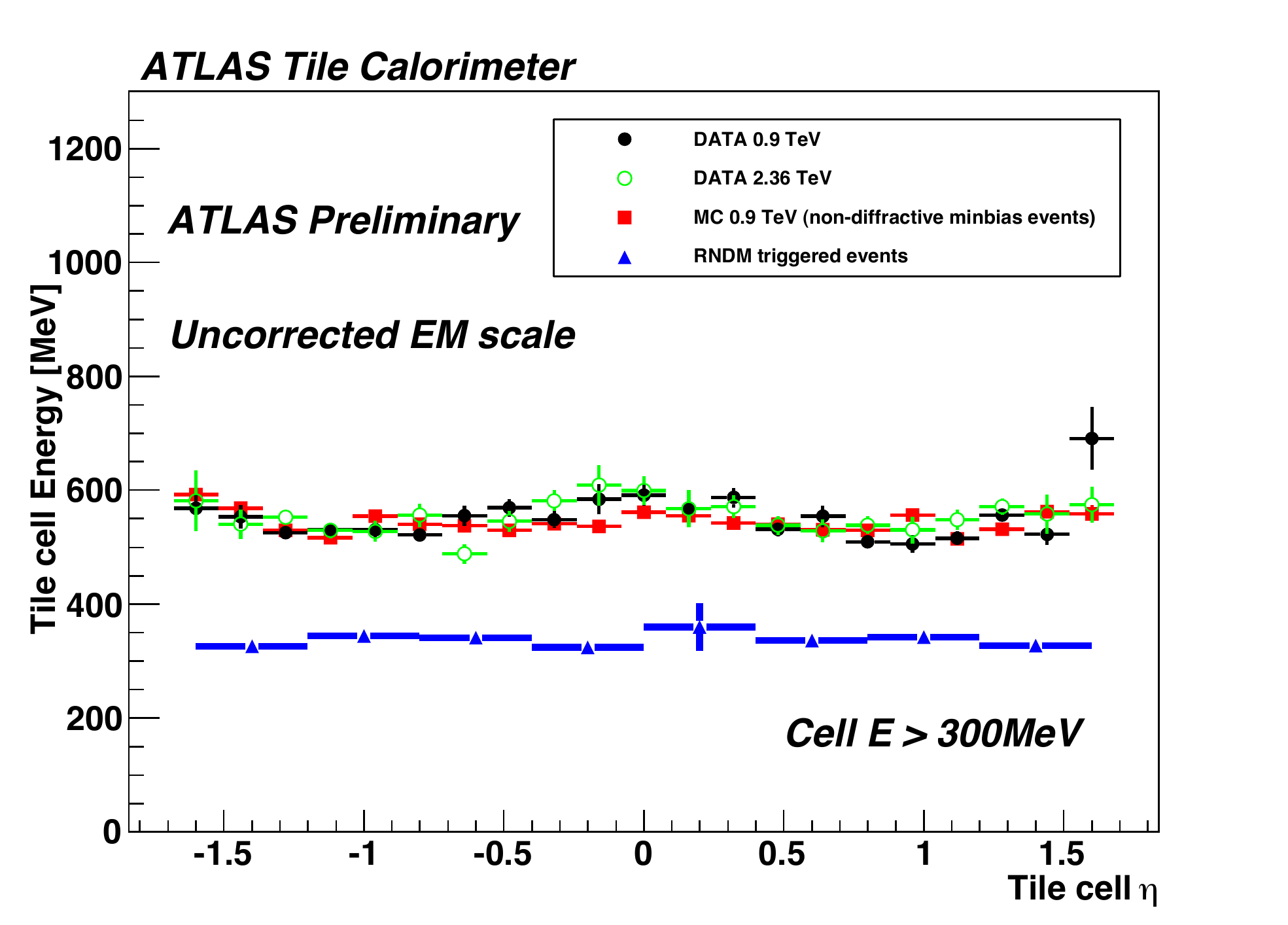}
\includegraphics[width=0.50\textwidth]{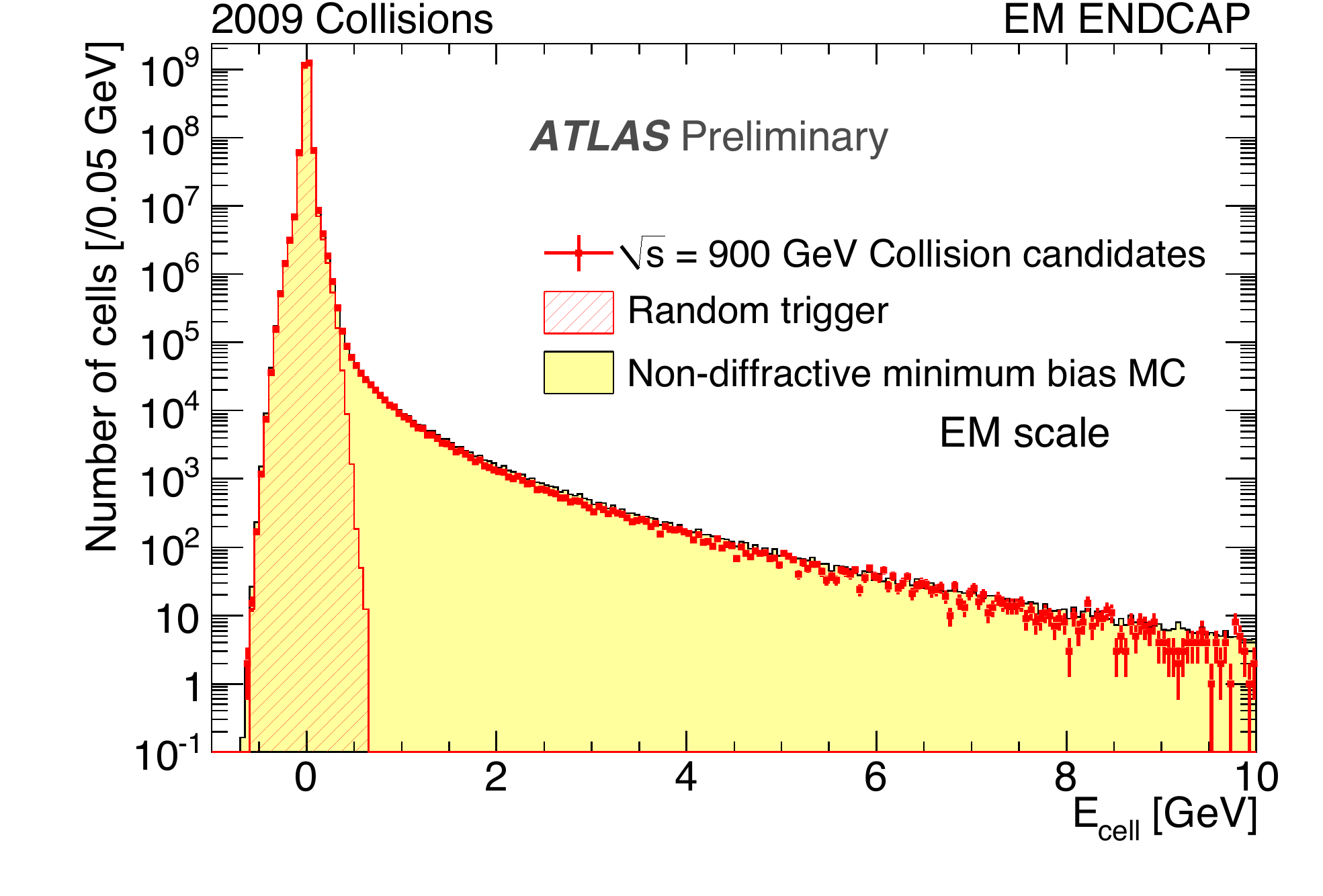}
\caption{Left: Average TileCal cell energy as a function of $\eta$ in collision events. 
Right: Distribution of cell energy with collision events for the LAr electromagnetic end-cap calorimeter (EMEC).   }
\label{fig-tilelar}
\end{figure}

The ATLAS calorimeter system consists of LAr sampling calorimeters for the electromagnetic barrel and end-cap calorimeters ($|\eta|<2.5$) as well as the hadronic end-cap ($1.5<|\eta|<3.2$) and forward ($3.2<|\eta|<4.9$) calorimeters. They are surrounded by an Iron/Scintillating Tile calorimeter consisting of a barrel section and extended barrel sections ($|\eta|<1.7$). The calorimeters have shown excellent performance, with very low and stable noise levels and uniform response already after the initial calibration cycles using cosmic ray and first collision data.

The average TileCal cell energy as a function of $\eta$ in collision events is shown in figure~\ref{fig-tilelar} (left). Only cells with energies above 300\,MeV are considered. Randomly triggered events with the same energy cut are superimposed with the collision candidate events and non-diffractive minimum bias Monte Carlo.
Figure~\ref{fig-tilelar} (right) shows the distribution of cell energy with collision events for the LAr electromagnetic end-cap calorimeter (EMEC). The cell energy distribution from random data (contribution of electronic noise only) is also shown, as well as non-diffractive minimum bias Monte-Carlo events. All cells, but ones known to be noisy and which are masked, are entered in the distributions. The distribution of cell energy in random events is not expected to be Gaussian, because the cell noise varies as a function of $\eta$. Similar performance has been observed in all parts of the calorimeter system.

\begin{figure}[t]
\includegraphics[width=0.50\textwidth]{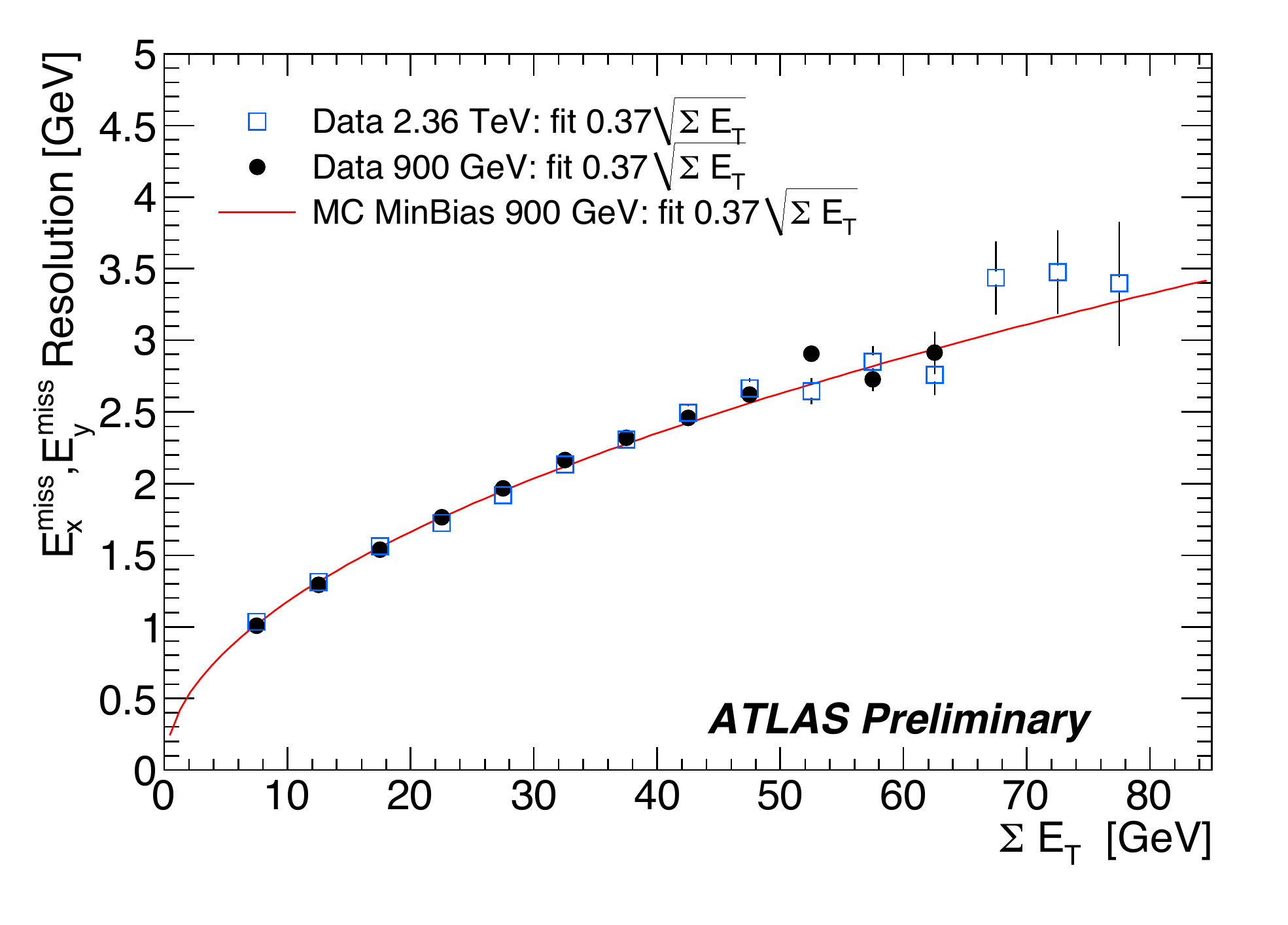}
\includegraphics[width=0.50\textwidth]{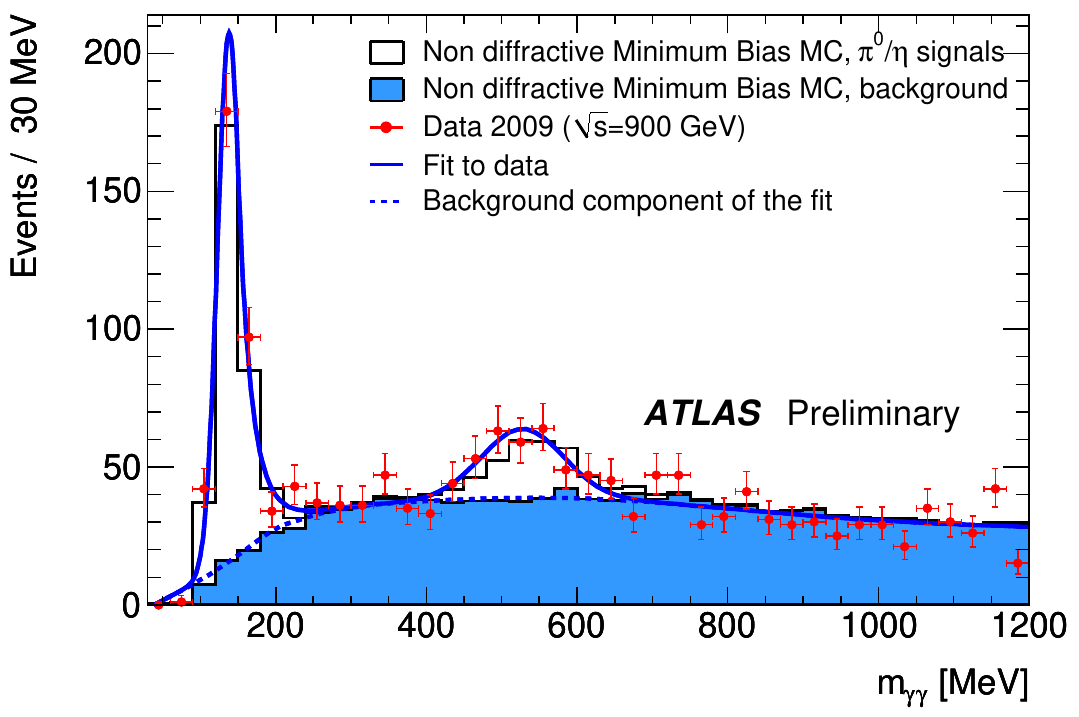}
\caption{Left:  $E_{\rm x}^{\rm miss}$ and $E_{\rm y}^{\rm miss}$ resolution as a function of the total measured transverse energy ($\sum E_{\rm T}$) for minimum bias events. The line represents a fit to the resolution obtained in the Monte Carlo simulation and the full dots (open squares) represent the results with data at 0.9\,(2.36)\,TeV. $E_{\rm x}^{\rm miss}$, $E_{\rm y}^{\rm miss}$, $\sum E_{\rm T}$ are computed with topocluster cells. 
Right: Di-photon invariant mass spectrum with tight selection cuts to extract the $\eta$ peak with the fit superimposed on the data. The Monte Carlo simulation sample is normalised to the number of entries in the data.}
\label{fig-metpi0}
\end{figure}

One of the most challenging calorimetric quantities to control especially in the early stages of data taking is missing energy, as it is subject to a range of possible influences from acceptance to inhomogeneous detector response. Figure~\ref{fig-metpi0} (left) shows the $E_{\rm x}^{\rm miss}$ and $E_{\rm y}^{\rm miss}$ resolution as a function of the total measured transverse energy ($\sum E_{\rm T}$) for minimum bias events. The line represents a fit to the resolution obtained in the Monte Carlo simulation and the full dots (open squares) represent the results with data at 0.9\,(2.36)\,TeV. $E_{\rm x}^{\rm miss}$, $E_{\rm y}^{\rm miss}$, $\sum E_{\rm T}$ are computed with cells clustered with a topological clustering algorithm using the electromagnetic energy scale. As for the tracking, known resonances can be used to study the performance of the calorimeters. Figure~\ref{fig-metpi0} (right) shows the di-photon invariant mass spectrum with tight selection cuts to extract the $\eta$ peak, with the fit superimposed to the data. The Monte Carlo simulation sample is normalised to the number of entries in the distribution for data. A less stringent selection has also been performed, and leads to a much higher statistics signal for the $\pi^0$ peak, also in agreement with expectations.

\section{Muon spectrometer}

The ATLAS Muon Spectrometer employs four detector technologies to provide both precision momentum measurement and trigger information for the barrel and end-cap regions. Monitored drift tubes provide the precise measurements in the bending planes of the barrel and end-cap regions, except in the very forward direction, which is covered by cathode strip chambers. The fast signals needed by the first level trigger are provided by resistive-plate chambers in the barrel, and thin-gap chambers in the end-cap region. The spectrometer is embedded in a toroidal magnetic field providing integrated bending power of between 2\,Tm and 5\,Tm, leading to an overall muon momentum resolution of less than 10\% for muons of up to about 1\,TeV energy. Muons may be reconstructed using information from the muon spectrometer alone, or using combined tracking between the muon spectrometer and the inner detector. Comparisons of the momentum measurement of muons matched between those two tracking devices allows to study the muon energy loss in the calorimeters. This becomes an important ingredient in the combined track reconstruction.

\begin{figure}[t]
\includegraphics[width=0.50\textwidth]{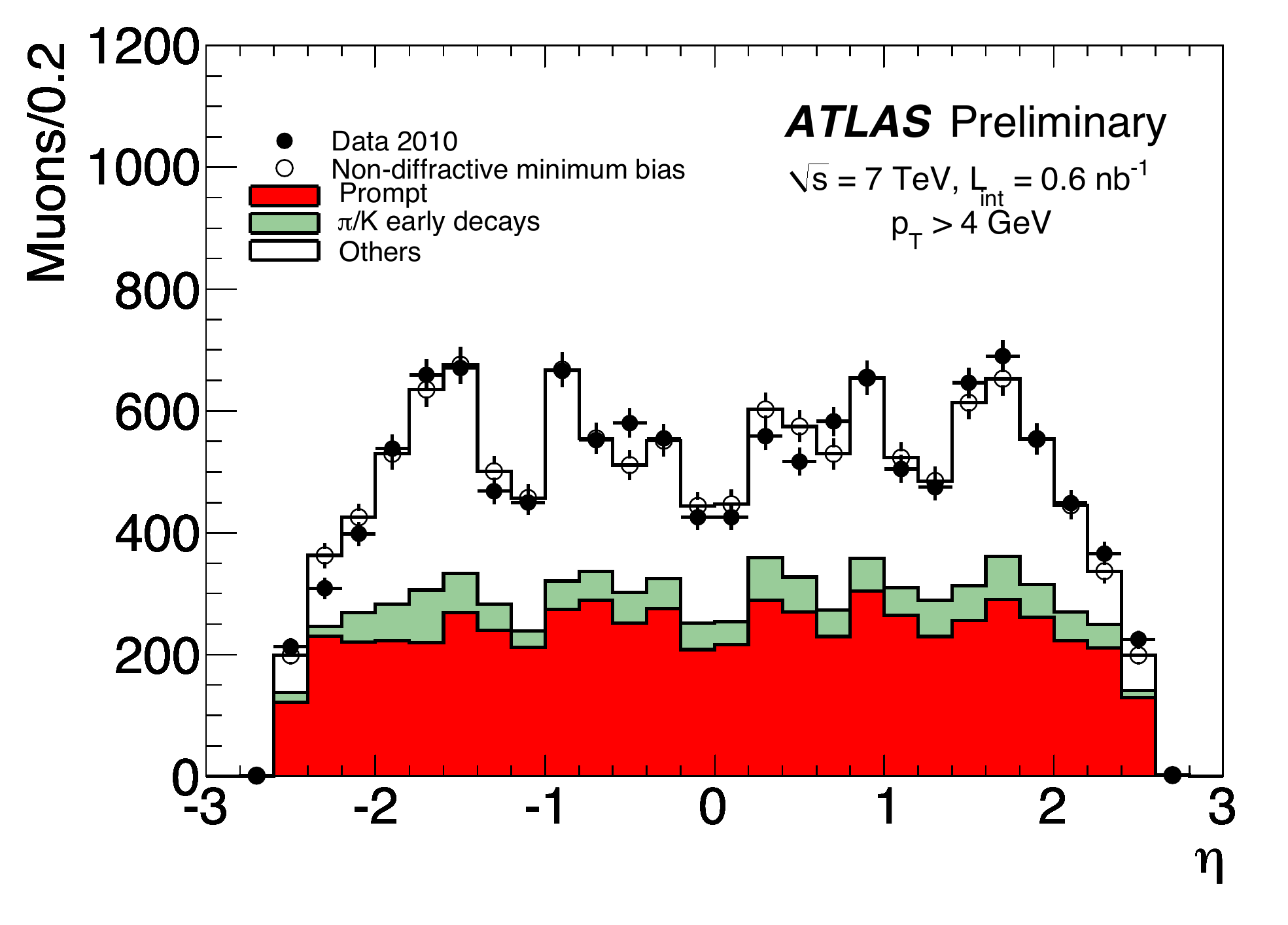}
\includegraphics[width=0.50\textwidth]{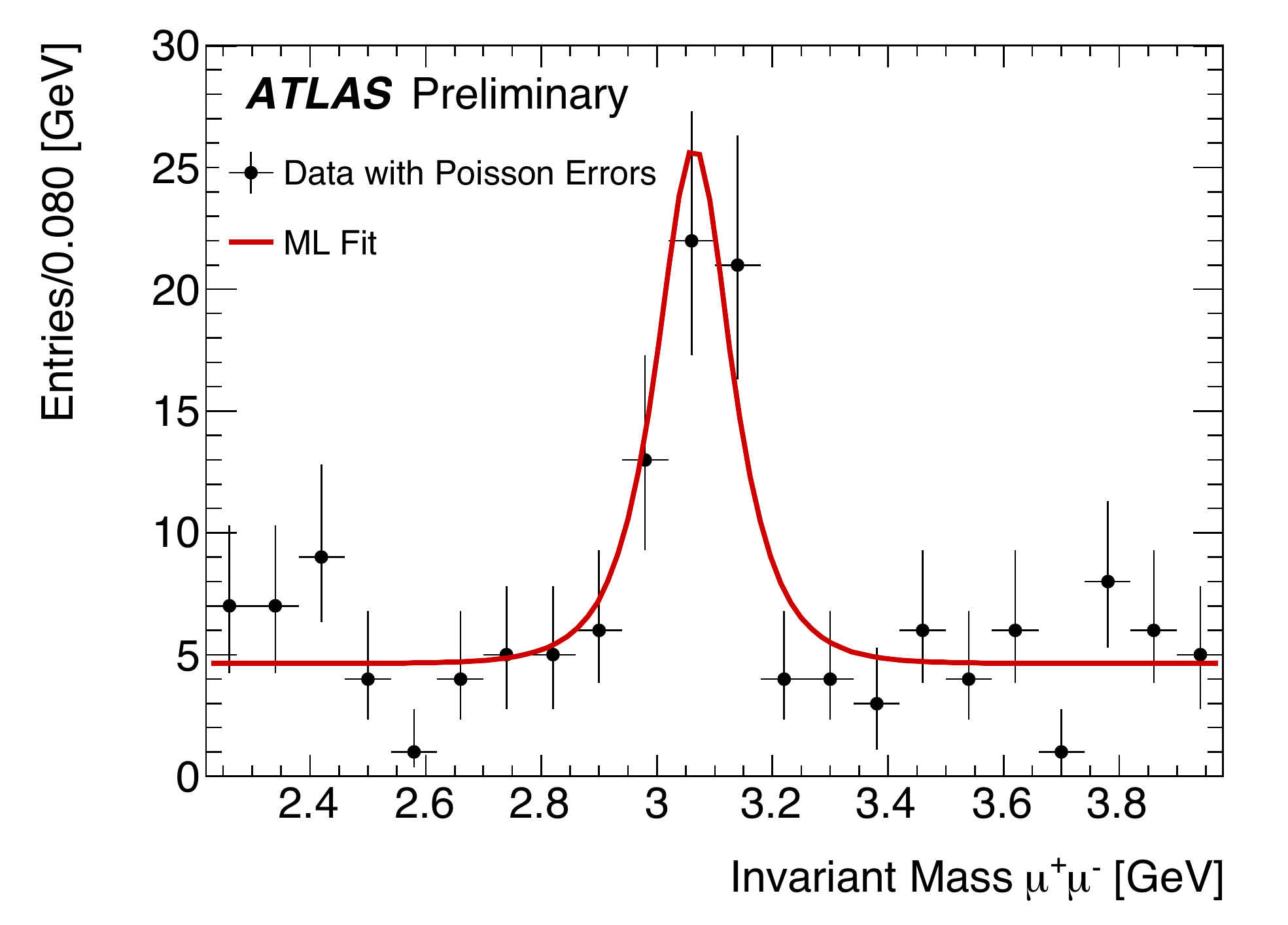}
\caption{Left: Muon pseudorapidity distribution for $p_{T}>4$\,GeV, normalised to the number of events.  
Right: Di-muon invariant mass distribution for oppositely charged muon pairs with an energy above 3\,GeV each.}
\label{fig-muon}
\end{figure}

Figure ~\ref{fig-muon} (left) shows the distribution in pseudorapidity of muons selected in 0.6\,nb$^{-1}$ of data at $\sqrt{s} = 7$\,TeV, with transverse momentum of more than 4\,GeV. The shape of the data distribution  is well reproduced by the Monte Carlo simulation, which has been normalised to the number of events found in the data.
In figure~\ref{fig-muon} (right) the di-muon invariant mass distribution is shown for oppositely charged muon pairs with an energy above 3\,GeV each. A clear $J/\Psi$ signal can be seen. The momentum is calculated from inner detector track parameters after a fit to a common vertex. At least one muon must be jointly reconstructed in the muon spectrometer and the inner tracking detector ('combined muon'). The fit to the distribution is an unbinned maximum likelihood fit, using event-by-event errors calculated for each di-muon. The mass region shown in the plot is fitted. The gaussian-mean mass is $3.06\pm 0.02$\,GeV, with a resolution of $0.08\pm 0.02$\,GeV, consistent with the $J/\Psi$. The number of signal events is determined to be $49\pm 12$, with the number of background events $28\pm 4$, both computed in a mass range 2.82 -- 3.30\,GeV (3$\sigma$ around the $J/\Psi$ peak). 

\section{Charged particle multiplicities}

The first published analysis result of the ATLAS collaboration reports on the measurement of inclusive charged particle distributions in $pp$ collisions~\cite{Aad:2010rd}. Measurements of these quantities have been carried out in the past in both $pp$ and $p\bar{p}$ collisions at a range of different centre-of-mass energies. In most cases, these measurements were obtained by selecting data with a double-arm coincidence trigger, with very limited acceptance for diffractive events. The data are then commonly corrected further to remove the remaining single-diffractive component, leading to what is called a non-single-diffractive (NSD) selection. The subtraction of diffractive components from the cross section involves model dependent corrections. In addition, one needs to correct for effects of the trigger selection on events with no charged particles within the acceptance of the detector. The measurement presented in~\cite{Aad:2010rd} uses a single-arm trigger over-lapping with the acceptance of the tracking volume. The results are presented as inclusive-inelastic distributions, with minimal model-dependence, since only one charged particle within the acceptance is required. Charged particles are required to have a momentum component transverse to the beam direction \pT\  > 500\,MeV, in the pseudorapidity range $|\eta|<2.5$. A total of 455,593 events were analysed in this study, corresponding to an integrated luminosity of approximately 9\,$\mu$b$^{-1}$ collected between December 6 and 15, 2009, at $\sqrt{s}=900$\,GeV. Cosmic ray events and beam induced background were studied as a possible source of contamination, and, for the selected sample, were found to be less then 10$^{-6}$ and 10$^{-4}$, respectively. The efficiency to trigger events in the selected sample was determined from data and is close to 100\%. The dominant systematic uncertainty of the measurement is associated to the track reconstruction efficiency, which was determined from Monte Carlo. After extensive studies of, among others, the effects of truth matching algorithms, misalignment effects and the impact of imperfections in the material description of the detector, an overall relative systematic uncertainty of 4.0\% was assigned to the track reconstruction efficiency for most of the kinematic range of this measurement, while 8.5\% and 6.9\% were assigned to the highest $|\eta|$ and the lowest \pT\ bins, respectively.

\begin{figure}
\includegraphics[width=0.50\textwidth]{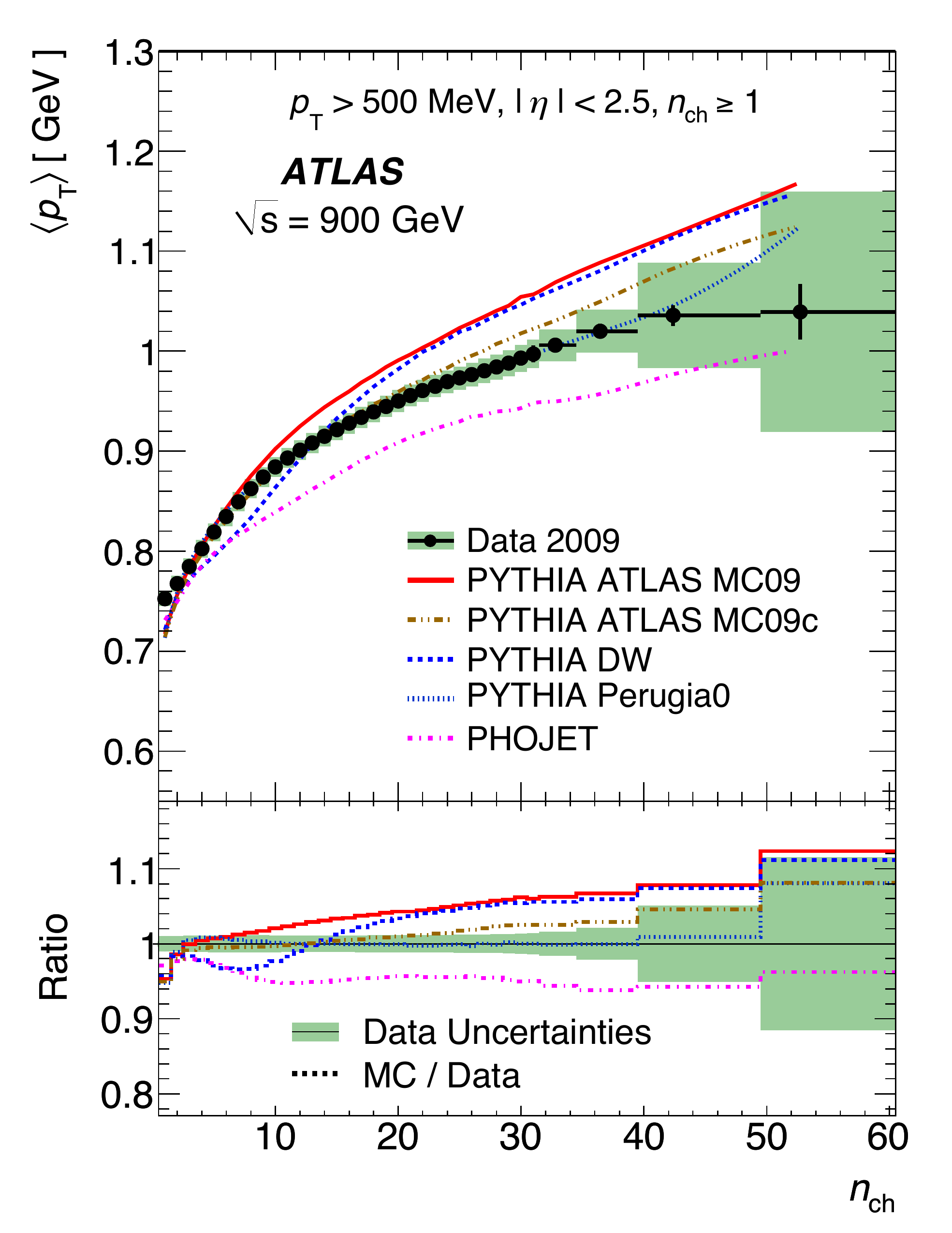}
\includegraphics[width=0.50\textwidth]{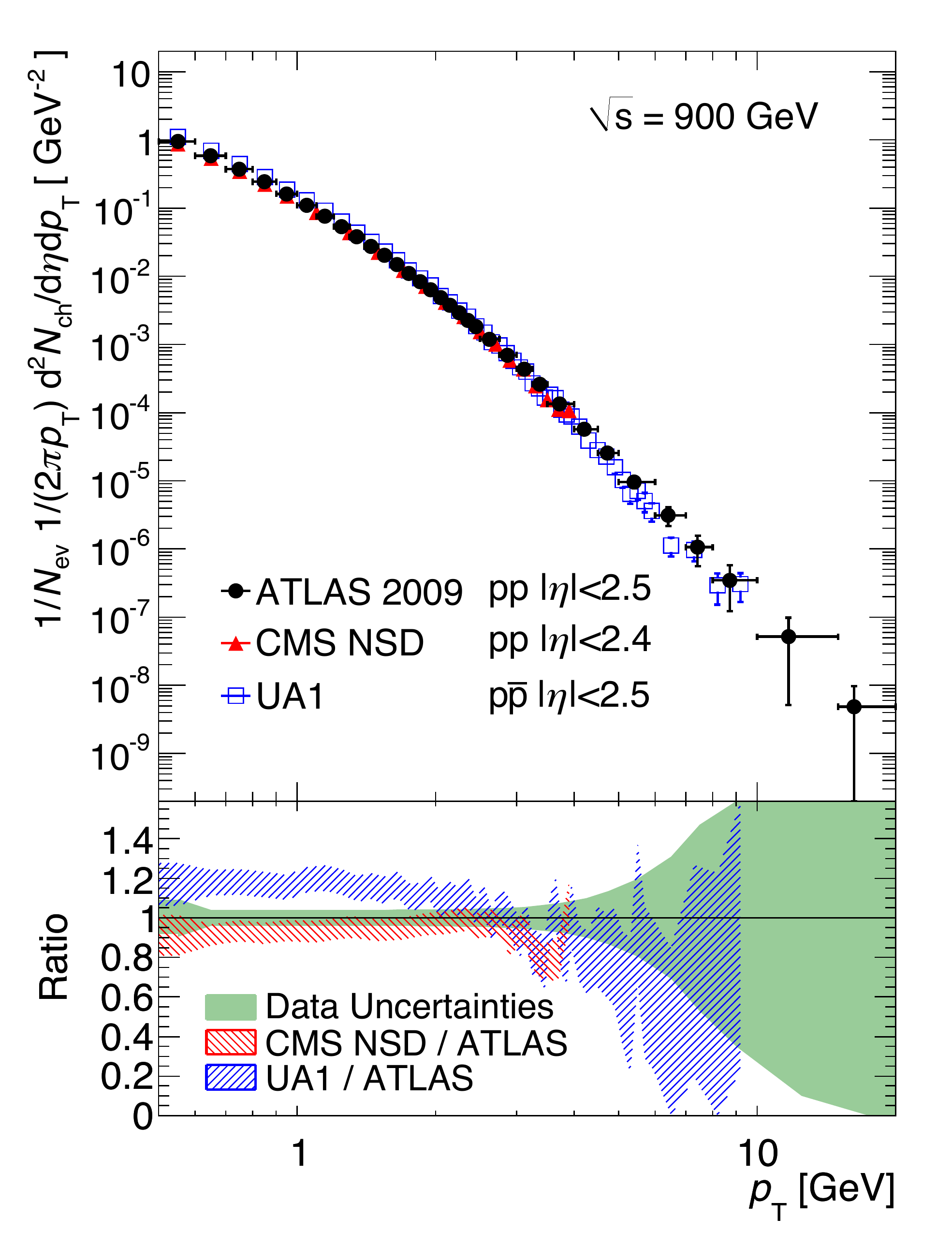}
\caption{Right: The average transverse momentum as a function of the number of charged particles in the event, for charged-particle multiplicities \nch\  > 1 within the kinematic range \pT \ > 500\,MeV and $|\eta| < 2.5$. The dots represent the data and the curves the predictions from different Monte Carlo models. The vertical bars represent the statistical uncertainties, while the shaded areas show statistical and systematic uncertainties added in quadrature. The values of the ratio histograms refer to the bin centroids. 
Left: The measured \pT\  spectrum of charged-particle multiplicities. The ATLAS $pp$ data (black dots) are compared to the UA1 $p\bar{p}$ data (blue open squares) and CMS NSD $pp$ data (red triangles) at the same centre-of-mass energy. }
\label{fig-nch}
\end{figure}

The distributions of  tracks reconstructed in the inner detector were corrected to obtain the particle-level distributions:
$$
\frac{1}{N_\mathrm{ev}}\cdot  \frac{\mathrm{d} N_\mathrm{ch}}{\mathrm{d} \eta}, \ \ \ 
\frac{1}{N_\mathrm{ev}}\cdot \frac{1}{2 \pi  p_\mathrm{T}} \cdot \frac{\mathrm{d}^2 N_\mathrm{ch}}{\mathrm{d} \eta \mathrm{d} p_\mathrm{T}}, \ \ \ 
\frac{1}{N_\mathrm{ev}}\cdot \frac{\mathrm{d} N_\mathrm{ev}}{\mathrm{d} n_\mathrm{ch}} \ \ \ 
{\rm and}
\ \ \ \langle p_\mathrm{T}\rangle ~ {\mathrm vs.} ~ n_\mathrm{ch}{\rm ,}
$$
where $N_\mathrm{ev}$ is the number of events with at least one charged particle inside the selected kinematic range, \Nch\ is the total number of charged particles, \nch\ is the number of charged particles in an event  and \meanpT\ is the average \pT\ for a given number of charged particles. 
Comparisons are made to previous measurements of charged-particle multiplicities in $pp$ and $p{\bar p}$ collisions at $\sqrt{s}=900$\,GeV centre-of-mass energies~\cite{Khachatryan:2010xs,Albajar:1989an} and to Monte Carlo models.

Figure~\ref{fig-nch} (left) shows the average \pT\ as a function of \nch. It can be seen to increase with increasing \nch\ and a change of slope is observed around \nch\ $=10$, a behaviour already observed in $p\bar{p}$ collisions at $\sqrt{s}=1.96$\,TeV by the CDF experiment~\cite{Aaltonen:2009ne}. The Perugia0~\cite{Sjostrand:2006za, Skands:2009zm} parameterization, which was tuned using CDF minimum-bias data at 1.96~TeV, describes the data well. The other models~\cite{Sjostrand:2006za,Albrow:2006rt}\cite{Engel:1994vs} fail to describe the data below \nch~$\approx 25$, with the exception of the PYTHIA-MC09c~\cite{Sjostrand:2006za,atlasmc09} tune, which optimises the the PYTHIA-MC09 tune also described in~\cite{atlasmc09} in the strength of the colour reconnection to  describe the \meanpT\  distributions as a function of \nch, as measured by CDF.

The \Nch\ distribution as a function of \pT\ in the kinematic range \pT\  $>500$\,MeV and $|\eta|<2.5$ is shown in figure~\ref{fig-nch} (right). The results of the CMS collaboration~\cite{Khachatryan:2010xs} for the same centre-of-mass energy are superimposed. As expected from the definition of NSD events in the CMS measurement, the number of charged particles in the CMS data is consistently lower than the data presented in this paper. For a more direct comparison the mean charged-particle density for ATLAS was recalculated in the range $\eta<2.4$ and a model dependent correction was applied to form an NSD particle density. The net effect of the correction is to reduce the charged-particle multiplicity, resulting in a value consistent with that published by CMS. Also overlaid on figure \ref{fig-nch} (right) are the UA1~\cite{Albajar:1989an} results, normalised by their associated cross section measurement. Compared to the present measurement they are approximately 20\% higher. Such a shift is expected from the double-arm scintillator trigger requirement used to collect the UA1 data, which excludes events with low charged-particle multiplicities.

\section{Conclusions}

The ATLAS experiment had a very successful start-up from the very first collisions provided by the LHC. The availability of the detector components and the data taking efficiency were close to 100\%. Many performance results had already been produced from the first data by the time of the conference, and even more have been produced since, showing a very good initial performance with considerable improvements based on systematic studies of collision data. The first ATLAS physics result had already been published by the time of the conference, measuring charged-particle multiplicities. With a value of $1.333\pm0.003\mathrm{(stat.)}\pm 0.040\mathrm{(syst.)}$ per event and unit of pseudorapidity at $\eta=0$ the result is 5-15\% higher than the Monte Carlo predictions.

\end{document}